\documentclass[epsfig,12pt]{article}
\usepackage{epsfig}
\usepackage{graphicx}
\usepackage{relsize}
\usepackage{amsmath,mathtools} 
\usepackage{amsfonts}
\usepackage{subfig}
\usepackage{color}
\usepackage{amssymb}
\usepackage{cite}
\usepackage{abstract}
\usepackage{tikz}
\usetikzlibrary{calc}
\usetikzlibrary{decorations.pathreplacing}

\newcommand{\tikzmark}[1]{\tikz[overlay,remember picture] \node (#1) {};}

\newcommand*{\BraceAmplitude}{0.5em}
\newcommand*{\VerticalOffset}{-2ex}
\newcommand*{\InsertOverBrace}[4][]{%
    \begin{tikzpicture}[overlay,remember picture]
\draw [decoration={brace,amplitude=\BraceAmplitude},decorate, thick,draw=blue,text=black,#1]
        ($(#2)+(0,-\VerticalOffset)$) -- 
        ($(#3)+(0,-\VerticalOffset)$)
        node [below=\VerticalOffset-2.5ex, midway] {#4};
    \end{tikzpicture}%
}%

\newcommand{\s}{\nobreak\hspace{.11em}\nobreak}
\newcommand{\beq}{\begin{equation}}   
\newcommand{\eeq}{\end{equation}}
\newcommand{\beqn}{\begin{eqnarray}}   
\newcommand{\eeqn}{\end{eqnarray}}

\newcommand{\Z}{\mathbb{Z}}
\newcommand{\jo}{\hspace{.08em}}
\newcommand\rank{\operatorname{rank}}
\newcommand{\me}{\mathrm{e}}

\newcommand{\gsim}{\lower.7ex\hbox{$
\;\stackrel{\textstyle>}{\sim}\;$}}
\newcommand{\lsim}{\lower.7ex\hbox{$
\;\stackrel{\textstyle<}{\sim}\;$}}
\renewcommand{\d}[1]{\ensuremath{\operatorname{d}\!{#1}}}

\usepackage{authblk}
\makeatother
\author[1]{Vladimir Bychkov\s}
\author[1,2]{Michael Kreshchuk\s}
\author[1]{Evgeniy Kurianovych\s}
\affil[1]{{\small
~School of Physics and Astronomy, University of Minnesota, Minneapolis, MN, 55455, USA}
}
\affil[2]{{\small
~Department of Physics, Tufts University, Medford, MA, 02155, USA
}}
\title{
{\Large{\textbf{
Strings and skyrmions on domain walls
}}}}
\date{}

\begin{document}
\maketitle
\begin{abstract}
We address a simple model allowing the existence of domain walls with orientational moduli localized on them. Within this model, we discuss an analytic solution and explore it in the context of previously known results. We discuss the existence of one-dimensional domain walls localized on two-dimensional ones, and construct the corresponding effective action. In the low-energy limit, which is the $\mathbb{O}(3)$ sigma-model, we discuss the existence of skyrmions localized on domain walls, and provide a solution for a skyrmion configuration, based on an analogy with instantons. We perform symmetry analysis of the initial model and of the low-energy theory on the domain wall world volume.    
\end{abstract}

\newpage

\section{Introduction}

In certain field theories, the existence of non-Abelian moduli localized on topological defects is possible. This fact was discovered originally for non-Abelian topological strings within the Super-Yang-Mills theories with matter~\cite{HananyTong,AuzziBolognesi,ShifmanYung-ConfMon}. Inspired by E.~Witten's work on cosmic strings~\cite{Witten}, a similar construction was developed for domain walls~\cite{Sh1,KurSh}. Also,~\cite{KurSh} proved the existence of such a construction numerically and analyzed its low-energy behaviour~--- the dynamics of the translational moduli of the domain wall and of the rotational non-Abelian moduli localized on the wall.

Here we expand the previously known results on non-Abelian moduli in several directions. First, we establish a correspondence between a recently found analytic solution~\cite{GaniLizRad, GaniLizRad2} for the domain wall profile, which supports non-Abelian moduli, and the results from~\cite{KurSh}. 
We also discuss the ways of using the analytic solutions as checks for numerical solutions. 
Then we explore a simplified model, the one with the $\mathbb{O}(3)$ symmetry reduced to $\Z_2$, and with only two scalar fields~--- the first of these creating a domain wall, and the other having a non-zero expectation value on the wall only. We find that the latter field itself can have a domain wall which, in the low-energy limit, behaves as a one-dimensional string-like object localized on a two-dimensional surface. We discuss its properties and low-energy dynamics, deriving a corresponding effective action. 

Within a construction which supports non-Abelian moduli described by an $\mathbb{O}(3)$ sigma-model, we discuss the possibility of existence of skyrmions of a type which is usually explored in condensed matter physics. We also give a solution for such a skyrmion, based on an analogy with instantons.

For both the initial model and the model describing non-Abelian moduli on the domain wall, we perform symmetry analysis, keeping in mind the possible applications of this method to more complex problems. The systems under consideration are very convenient for testing this method, since they allow us to compare the ensuing results to the known answers. A review of the symmetry analysis method is provided.

The paper is organized as follows. Section~\ref{Moduli on a domain wall} reviews construction of moduli on domain walls. Section~\ref{Analytic solution} discusses an analytic solution of such a system, the conditions of its existence, and its properties. We also compare this solution with the results from~\cite{KurSh}. Section~\ref{1d wall on a 2d wall} describes a one-dimensional wall localized on a two-dimensional one. In Section~\ref{SectionSkyrmions}, a skyrmion localized on a domain wall is addressed. In Sections~\ref{Moduli on a domain wall} and~\ref{SectionSkyrmions}, we provide the symmetry analysis of the systems under consideration. The mathematical tools for this analysis are reviewed in the Appendix.

\section{Moduli on a domain wall}  \label{Moduli on a domain wall}

To review the construction of moduli on a domain wall~\cite{Sh1,KurSh}, we start out with the Lagrangian of a real scalar field:
\beq
{\cal L}_0=\frac{1}{2}\partial_{\mu}\varphi\partial^{\s \mu}\varphi-V(\varphi)\quad, 
\qquad
V(\varphi)=\lambda(\varphi^{2}-v^{2})^{2}
\quad.
\label{L0}
\eeq
It has a $\Z_2$ symmetry ($\varphi\to  -\varphi$) which is spontaneously broken in a vacuum~--- so $\varphi$ assumes the value of either $v$ or $-v$. This model supports domain walls created in the region connecting the two vacua. Let us assume that a wall is parallel to the $(x,y)$ plane and that $\varphi=\pm v$ for $z=\mp \infty$. Hence, the solution for $\varphi$ can be found analytically:
\beq
\varphi(z)=-v\tanh\left[\frac{m_{\varphi}}{2}(z-z_{0})\right]\quad,
\label{tanh}
\eeq
where $m_{\varphi}=\sqrt{8\lambda v^{2}}$ is the mass of the $\varphi$  field and $z_{0}$ is the wall centre.

Now, let us add to the above model a triplet of fields $\chi^{i}$, $i=1,2,3$, described by the Lagrangian
\beqn
{\cal L}_{\chi}&=&\frac{1}{2}\partial_{\mu}\chi^{i}\partial^{\s \mu}\chi^{i}-U(\chi,\varphi)\quad,\qquad
\label{Lchi}
\nonumber\\
U(\varphi,\chi^i)&=&\gamma\left[(\varphi^{2}-\mu^{2})\chi^{i}\chi^{i}
+\beta(\chi^{i}\chi^{i})^{2}\right]\quad, \qquad v^2>\mu^2\quad,
\label{U}
\eeqn
so that the new model has the Lagrangian ${\cal L}={\cal L}_0+{\cal L}_{\chi}$. Then, for a choice of the parameters\s\footnote{~For more details about constraints on the set of parameters and stability of the solution under consideration see~\cite{KurSh,GaniLizRad,MShYu}.} for which the vacua are given by $\varphi^2 = v^2, \chi^i = 0$, we still can have a domain wall, connecting these two vacua, and $\chi^i$, getting a non-zero expectation value only inside of that wall. Since some energy in the wall is taken by the kinetic term of $\chi^i$ and still the total energy should be smaller than without $\chi^i$ condensation, the kinetic term of $\varphi$ should be smaller and, therefore, the transition region between vacua (i.e. the wall) should become wider.

To get a domain-wall solution explicitly, we need the Euler-Lagrange equations following from ${\cal L}$:

\beq
\left\{\begin{alignedat}{9}
\varphi''&=4\lambda\varphi(\varphi^{2}-v^{2})&&+2\gamma\chi^{2}\varphi \\
\chi^{i\,\prime\prime}&=2\gamma(\varphi^{2}-\mu^{2})\chi^{i}&&+4\beta\gamma\chi^{2}\chi^{i}
\end{alignedat}\right.
\quad.
\label{system}
\eeq
\begin{sloppypar}
The solution minimising the domain wall energy is degenerate and still has the $\mathbb{O}(3)$ target space symmetry of the Lagrangian (\ref{U}). This symmetry has to be global, and $\chi^i$ should point in the same target-space direction in all points of the real space. Indeed, if we allow target space rotations that depend on the coordinates along the domain wall plane, this will create additional kinetic terms arising from $x$ and $y$ derivatives and, therefore, will increase the total energy. So, we can look for the solution for $\chi^i$ pointing in the same (arbitrary) direction everywhere in the target space.
For example, let us leave only the $\chi^3$ component non-zero and denote it with $\chi$. The corresponding system of Euler-Lagrange equations will be given by expressions~\eqref{system} with ${\chi^{i} = \{0,0,\chi\}}$ plugged therein, and we shall also refer to it as~\eqref{system}.
This system has the first integral
\end{sloppypar}
\beq
\frac{1}{2}\left(\frac{\d\varphi}{\d z}\right)^{2}+\frac{1}{2}\left(\frac{\d\chi}{\d z}\right)^{2} - U\left[\varphi(z),\chi(z)\right] - V[\varphi(z),\chi(z)]\quad.
\label{FirstIntegral}
\eeq

The solution of (\ref{system}) with a domain wall created by $\varphi$, and with $\chi$ localized on it, was obtained in~\cite{KurSh}. After an energy-minimising configuration is obtained, we can restore the $\mathbb{O}(3)$ symmetry of the $\chi$ field, allowing its position in target space to depend on the space-time coordinates on the wall, while $\chi^i \chi^i$ is still determined by the minimal-energy condition and is therefore localized on the $\mathbb{S}^2$ sphere in the target space. This then gives birth to non-Abelian moduli on the domain wall. The corresponding effective Lagrangian was derived in~\cite{KurSh}.

In this paper, we consider a relatively simple Lagrangian which allows us to discover easily the symmetries of the system and the moduli which emerge after breaking of some of these symmetries. For more complex Lagrangians, the ensuing symmetries may be harder to observe.  Systems which do not allow a Lagrangian description may be considered as well. In such situations, a general mathematical technique which allows to analyze symmetries of differential equations and integrate them using those symmetries may be a very powerful tool. Here and hereafter in our analysis of skyrmions, we shall perform such an analysis for the purpose of checking our results and in quest for symmetries which may be unnoticed by initial investigation. This will provide a demonstration of the general method of symmetry analysis on a simple and well-understood set-up.

For details of the said general method, we refer the reader to the Appendix below and to the book by Peter Olver~\cite{Olver}. For heavy-duty calculations that are often needed in the symmetry analysis, we recommend the software provided with the book~\cite{Lie}.

In the case of system~\eqref{system}, the symmetry analysis renders only the translational symmetry along the $z$-direction. So there are no other symmetries that we might had missed in our discussion. A search for generalized symmetries furnishes no new results either.

\section{Analytic solution}  \label{Analytic solution}

The numerical method developed in~\cite{KurSh} can provide a solution for any set of parameters for which such a solution exists. Still, it is worth looking for an analytic solution of the same problem, since it can be easier to analyze qualitatively. However, finding such a solution may result in certain loss of generality, since its existence can impose additional constraints on the set of parameters.

Following~\cite{GaniLizRad}, let us consider the solution for the $\varphi$ field in the form (\ref{tanh}) which $\varphi$ acquires in the absence of~$\chi^{\s i}$:
\beq
\varphi(z) = - v\tanh(\alpha z)\quad.
\label{anphi}
\eeq
 Here we placed the centre of the wall at $z_0 = 0$. The coefficient $\alpha$, which determines the  width of the wall, is to be fixed later. Then, from the first equation in (\ref{system}), we obtain:
\beq
\chi^2 = \frac{2\lambda v^2 - \alpha^2}{\gamma} \frac{1}{\cosh^2 \alpha z} \equiv
\frac{A^2}{\cosh^2 \alpha z}\quad.
\label{anchi}
\eeq

Substituting this expression into the second equation of (\ref{system}), we end up with a polynomial function of $(\cosh \alpha z)^{-1}$. Equating all of its coefficients to zero yields
\begin{alignat}{9}
\alpha^2 &= 2\gamma(v^2 - \mu^2)\quad&&,
\label{alpha}
\intertext{and}
\alpha^2 &= \gamma(v^2 - 2\beta A^2)\quad&&.
\label{constraint}
\end{alignat}

So, for a field $\varphi$ creating a domain wall and a field $\chi$ localized on it, we have obtained an analytic solution given by (\ref{anphi}) and (\ref{anchi}), respectively. The value of $\alpha$ is given by (\ref{alpha}). Since $\alpha$ and $A$ are functions of the parameters of the model, (\ref{constraint}) gives a constraint on a set of parameters, under which such a solution exists.

Note that for $\varphi$ and $\chi$ given by the above formulae the following equality holds:
\beq
\frac{\varphi^2}{v^2} + \frac{\chi^2}{A^2} = 1\quad.
\label{ellypse}
\eeq
We see that the solution corresponds to an ellipse in the ${(\varphi,\chi)}$ plane. In this plane, the vacua correspond to the ends of the major axis where ${\varphi = \pm v}$ and ${\chi = 0}$.

It is worth noting that the solution~\eqref{anphi} for $\varphi$ has the same form as~\eqref{tanh}, which was obtained without condensing of $\chi$. However, in this case the width $\alpha^{-1}$ of the wall created by $\varphi$ is determined by the parameters $\gamma$ and $\mu$ from the $\chi$ part of the Lagrangian. The derivation above is inverse to the one provided in~\cite{Rajaraman}, where a similar result is produced starting from the assumption that the solution in $(\varphi,\chi)$ plane has an elliptic shape.

It would be interesting to compare this analytic result with the numerical procedure applied in~\cite{KurSh}. Such a comparison may help to check the precision of the numerical solution. Mind though that this comparison cannot be carried out for the set of parameters used in~\cite{KurSh} for they do not satisfy~\eqref{constraint}.

The main difficulty in calculating numerically the domain wall configuration is that the boundary conditions are given only at the infinity and correspond to the equilibrium state. Therefore, if we solve the differential equation with these conditions, we shall always remain in the same state. To evade this difficulty, the following approach was used in~\cite{KurSh}: both fields were slightly perturbed away from their equilibrium values (equal to the values at infinity). Thereafter, the shooting method was used: the initial deflection was varied until it gave the expected results near the other infinity. 

\begin{sloppypar}
This approach required the knowledge of the asymptotics near infinity. Those were obtained from (\ref{system}) by keeping only the terms of the first order in $\varphi$ and $\chi$. However, for the solution (\ref{anphi},~\ref{anchi}) this approach fails since the variation of $\varphi$ from the vacuum ${\eta \equiv v - \varphi}$ will be of the order of $\chi^2$, as can be seen from (\ref{ellypse}). So the simple asymptotics of~\cite{KurSh} do not work. Of course, it is possible to obtain the asymptotics directly from the analytic solution~\cite{GaniLizRad}, but this will be a separate calculational procedure. When we have neither an analytic solution nor the ability to simplify (\ref{system}), a configuration can be found only through a more complicated numerical procedure, involving variation of both the fields and their derivatives.
\end{sloppypar}

A different strategy of solving system~\eqref{system} numerically may be of help, though. The form of the analytic solution (\ref{anphi},~\ref{anchi}) suggests such a change of variables that the new variable is defined on a finite interval. Under this kind of  transformation, the vacuum values of the fields at infinity become the boundary values at the ends of the said finite interval. Specifically,~\eqref{system} compels us to try the change of variables
\footnote{~For this purpose, one can use any function mapping an infinite interval onto a finite segment, e.g., $z = \kappa \tan \dfrac{\xi \, \pi}{2}$ or $z = \dfrac{\xi}{(1-\xi)}\;$.}
\begin{equation}
    z\, = \kappa\,\arctan \xi \quad,
    \label{chacha1}
\end{equation}
with $\kappa$ being a free parameter, and $\xi$ being a new argument defined on the interval ${[-1, 1]}$. However, under this substitution, the equations acquire singular points at the ends of the interval, even though the desired solutions stay nonsingular there.
Having performed a substitution~\eqref{chacha1}~---~or another transformation to a variable belonging to a finite domain~---~one can choose from an arsenal of methods developed to solve the boundary values problem.

In~\cite{GaniLizRad}, an important problem of energy levels of a ``bare'' domain wall (with $\chi = 0$ everywhere) was discussed. Based on the known spectrum of a modified P\"{o}schl-Teller potential, it provided a much better estimate than~\cite{KurSh} where the potential was approximated with a parabolic well. In general, neither of these results renders exact values of the energy levels of a non-``bare'' wall, because the condensation of $\chi$ also changes the profile of $\varphi$.

\section{1d wall on a 2d wall}  \label{1d wall on a 2d wall}

The Lagrangian with a fixed orientation of $\chi$ in the target space can be generalized, allowing $\chi$ to have negative values as well as positive:
\beq
{\cal L}_{\Z_2} = \frac{1}{2} (\partial \varphi)^2 + 
\frac{1}{2} (\partial \chi)^2 -
\lambda(\varphi^{2}-v^{2})^{2}
- \gamma\left[(\varphi^{2}-\mu^{2})\chi^2 +\beta\chi^4 \right]\quad.
\label{Z2Lagrangian}
\eeq
This way we obtain a model with just two real scalar fields $\varphi$ and $\chi$; its relevance to the triplet model above will be discussed at the end of this chapter.

Note that this Lagrangian possesses a $\Z_2$ symmetry, ${\chi \rightarrow - \chi}$. Consequently, if a configuration of the fields $\varphi$ and $\chi$ satisfies the equations of motion (\ref{system}), then a configuration with $\varphi^{\,\prime}=\varphi$ and $\chi^{\,\prime}=-\chi$ satisfies (\ref{system}) as well. Of course, this holds for the construction described above: if we have a domain wall created by the field $\varphi$ and the field $\chi$ localized on it, then the solution for $\chi$ with the opposite sign can exist as well and will have exactly the same energy. Visible from (\ref{anchi}), where $\chi$ is defined up to a sign, this fact also holds in a more general case where we do not have an analytic solution. A static configuration on the wall breaks this $\Z_2$ symmetry, picking one of the two configurations with the same energy, and this allows for emergence of topological defects.

Let us assume that the $\varphi$ field creates a domain wall in the ${(x,y)}$ plane, with the centre of the wall residing at $z = 0$. Then, assuming that at ${y = - \infty}$ we have a solution with positive $\chi$, and for ${y = + \infty}$ with negative $\chi$, we acquire a transition region~--- a domain wall for $\chi$, localized on a domain wall of $\varphi$. We'll also call this one-dimensional domain wall a domain line. Let's also assume that the transition region is centerd at $y=0$. A similar construction was considered in~\cite{Auzzi}.

Clearly, the profile of $\varphi$ in the domain line will be different from the profile of $\varphi$ on the wall far away from the line, i.e., $\varphi(x,0,z)\neq\varphi(x,\pm\infty,z)$ when $z$ is near $0$. This happens since, as we have demonstrated above, a non-zero expectation value of $\chi$  makes the transition region for $\varphi$ along $z$  wider. However in the domain line $\chi$ is close to zero, so the transition region for $\varphi$ is narrower (see also Fig.~3 in~\cite{KurSh}).

At a larger scale, the wall of $\varphi$ can be regarded as a two-dimensional surface, and the wall of $\chi$~--- as a one-dimensional string localized on it. Let us take a closer look at the low-energy translational moduli of this string. In the low-energy limit, the profile of the domain line can be considered unchanged, but its center can experience transversal displacements depending on $x$ and time:
\beq
y=y(x,t)\quad, \qquad z=z(x,t)\quad.
\eeq
Substituting these equations into the Lagrangian ${\cal L}={\cal L}_0+{\cal L}_{\chi}$ gives the following equation for $\chi$:
\beq
\begin{multlined}
(\partial \chi)^2 = 
- \left(\frac{\partial \chi}{\partial y}\right)^2 - \left(\frac{\partial \chi}{\partial z}\right)^2
- \left(\frac{\partial \chi}{\partial y}\right)^2 \partial_q y_0 \partial^{\s q} y_0
- \left(\frac{\partial \chi}{\partial z}\right)^2 \partial_q z_0 \partial^{\s q} z_0\quad,
\label{StringKinetic}
\end{multlined}
\eeq
and an analogous expression for $\varphi$. Here we only kept the  relevant terms (the integral of the potential term will remain unchanged, since the profiles of the domain walls are the same). Omitting the inessential constants, arising after the integration of the first two terms in (\ref{StringKinetic}), we obtain the following effective string action:
\beq
\Delta S = \int \d t \d x \left[B_y (\partial_q y_0 \partial^{\s q} y_0) + B_z (\partial_q z_0 \partial^{\s q} z_0) \right]\quad,
\label{StringAction}
\eeq
where
\begin{subequations}
\begin{alignat}{9}
B_y &= - \int \d y \d z \left[\left(\frac{\partial \chi}{\partial y}\right)^2 +
\left(\frac{\partial \varphi}{\partial y}\right)^2 \right]\quad&&,
\\
B_z &= - \int \d y \d z \left[\left(\frac{\partial \chi}{\partial z}\right)^2 +
\left(\frac{\partial \varphi}{\partial z}\right)^2 \right]\quad&&.
\end{alignat}
\end{subequations}
The second term in (\ref{StringAction}) should be taken into account due to the fact noted above: the presence of a domain line of $\chi$ changes the profile of the two-dimensional wall of $\varphi$. So the effective action here is also different from the low-energy domain wall action, derived in~\cite{KurSh}. If we are interested in an effective action of a string on a wall, then only the first term in (\ref{StringAction}) should be considered.

Let us explore the conditions of existence of the domain line solution described above, and let us study its stability. There are two different aspects of the problem. First of all, there should exist a solution which makes the condensation of the $\chi$ field inside a domain wall energetically advantageous and stable. The  existence of such a solution, along with the stability condition, impose constraints on the parameters of the model. Such constraints were extensively investigated in~\cite{KurSh,GaniLizRad,MShYu}, and that analysis is applicable to our setting. This applicability stems from the fact that the  energy-minimizing solution with the $\chi$ condensation defines only a profile of $\chi(z)^2$ but does not affect any structure that $\chi$ can have. The argument holds, in particular, when $\chi$ is an $\mathbb{O}(3)$ triplet of fields, $\chi^2=\chi^i \chi^i, i=1,2,3\s$, as was considered above. It also holds for a $\Z_2$-symmetric singlet, a case in which we are interested in this section. It will also work for any other model with a symmetry allowing $\chi^2$ to be fixed.

Suppose the constraints in the space of parameters are imposed. Then, for the $\Z_2$ symmetric model (\ref{Z2Lagrangian}) we are considering here, the stability of a domain line is established on topological grounds (see e.g.~\cite{ShBook}). We have a topologically non-trivial mapping of the real space boundary, ${y=\pm\infty}$, into the vacuum manifold. Such a mapping cannot be deformed into a topologically trivial one without making the energy of the system infinite in the process. 

The stability of the moduli configuration was proved in~\cite{KurSh}.  The stability of the domain-line solution within an effective field theory on a domain wall is guaranteed by topological arguments. Now, what if such a domain-line solution, albeit stable within the effective field theory, is unstable within the general theory with the initial Lagrangian~\eqref{Z2Lagrangian}? If a domain-line defect is unstable, it should have an energetically favourable mode of decay. The only possibility for the decay would be the region with $\chi=0$ in the domain-line core spreading infinitely wide and leaving no $\chi$ condensate on the wall.  This, however, contradicts the fact (proved by~\cite{KurSh}), that the solution with non-zero $\chi$ is energetically favourable. Therefore, the domain line must be stable.

\section{Skyrmions}  \label{SectionSkyrmions}

\begin{figure}
      \epsfxsize=226px
   \centerline{\epsffile{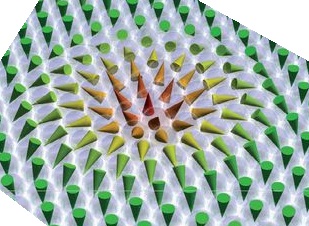}}
   \caption{The simplest skyrmion: target space vector goes from one pole at the centre to another one at infinity, while its projection to the plane always remains parallel to the radius-vector. Image from~\cite{RommingHanneken}.}
   \label{hedgehog}
\end{figure}

\begin{figure}
      \epsfxsize=217px
   \centerline{\epsffile{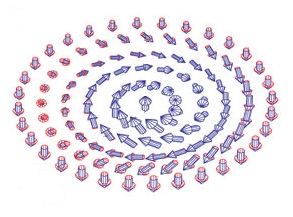}}
   \caption{Spiral skyrmion: target space vector winds around while going from one pole at the centre of the defect to another one at infinity. Image from~\cite{RosslerBogdanov}.}
   \label{spiral}
\end{figure}

\begin{figure}
\centering
  \subfloat[$S^1$.]{%
    \includegraphics[width=0.5\textwidth]{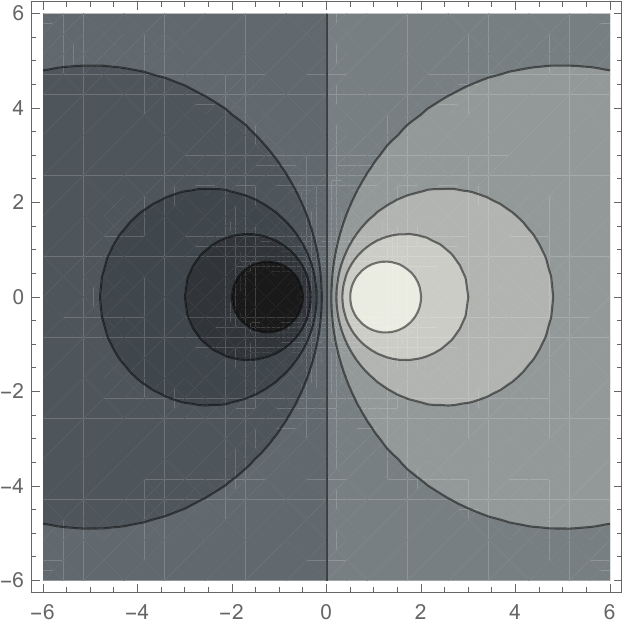}}
  \subfloat[$S^2$.]{%
    \includegraphics[width=0.5\textwidth]{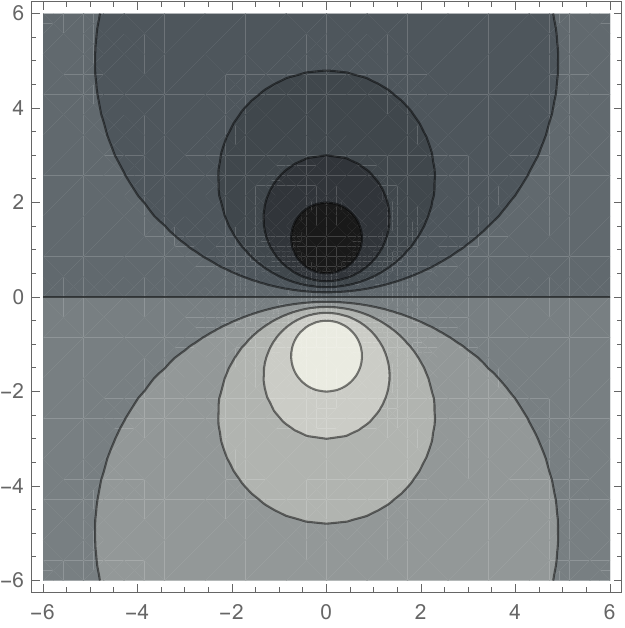}}
    \includegraphics[width=0.09\textwidth]{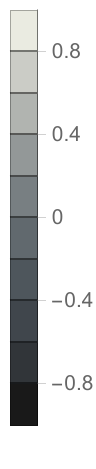} \\
  \subfloat[$S^3$.]{%
    \includegraphics[width=0.5\textwidth]{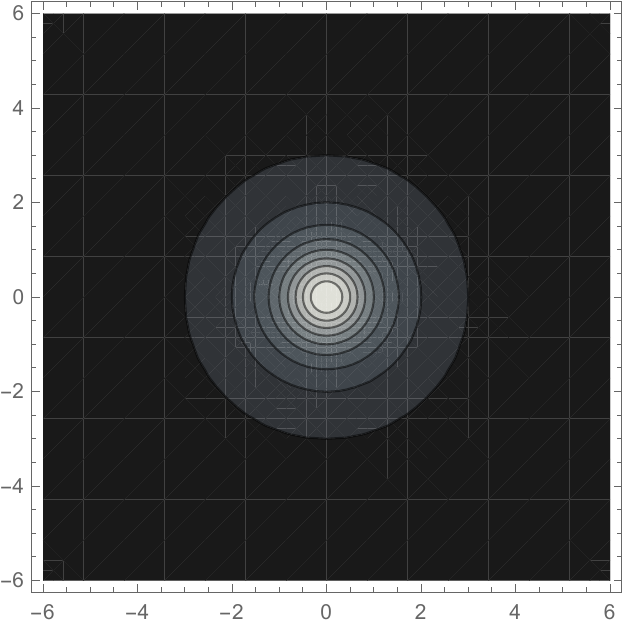}} 
  \subfloat[$S$ as a vector on the domain wall plane.]{%
    \includegraphics[trim={0 3cm 0 3cm},clip,width=0.7\textwidth]{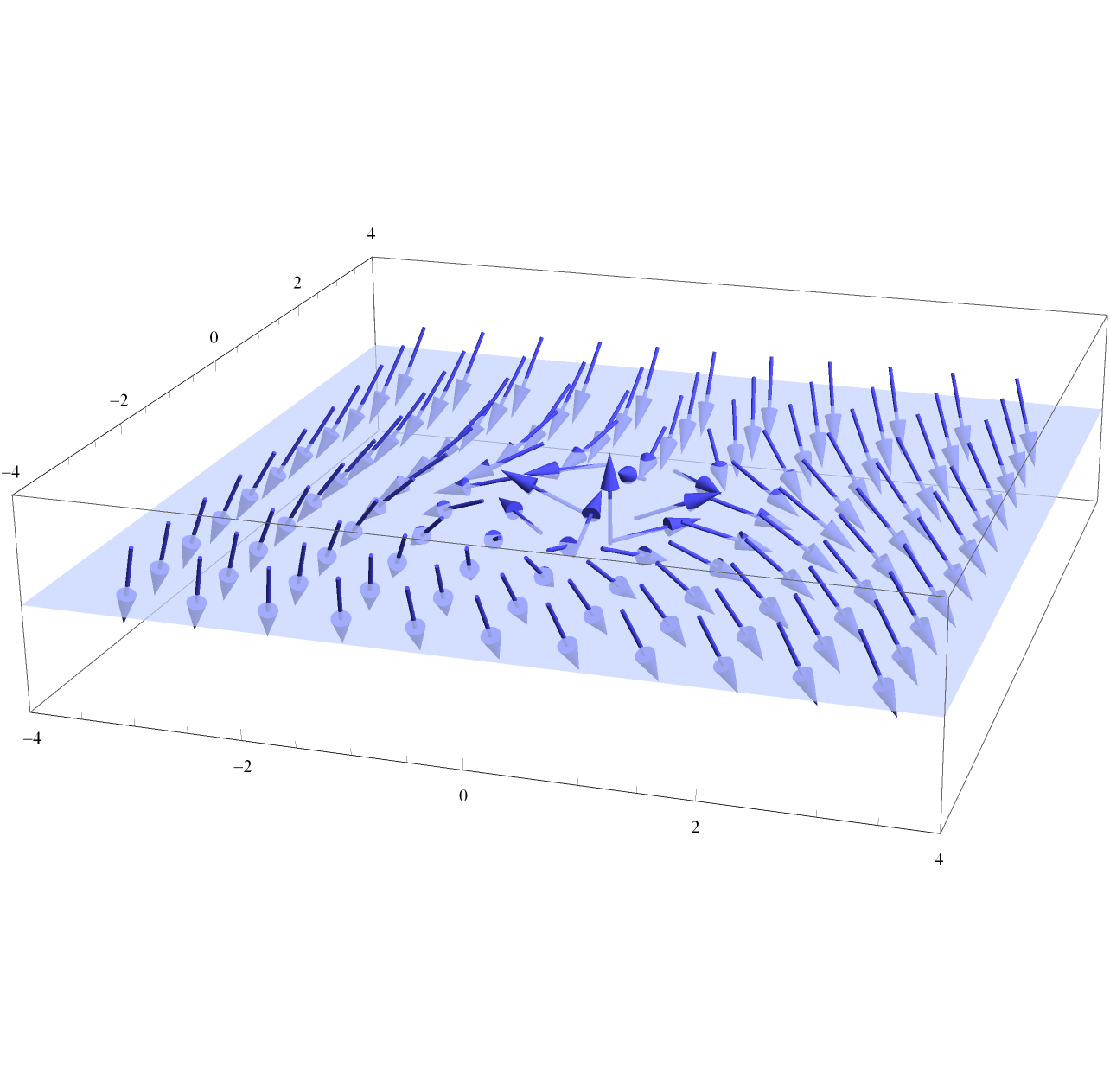}}
  \caption{(a~--~c) --- components of $S$ from (\ref{SkyrmionSolution}) as functions of coordinates on the domain wall plane, measured in the units of $\vert\vec{y}\vert$, (d) --- $S$, visualized as a three-dimensional unit vector on the domain wall plane, whose $x$, $y$ and $z$ components are $S^1$, $S^2$ and $S^3$ respectively.}
  \label{SkyrmionImages}
\end{figure}

Now, let us restore the $\mathbb{O}(3)$ symmetry of the $\chi$ field. The magnitude of the vector $\vec{\chi}$ in the target space in each point of the real space is determined by energy minimisation. However, it has a freedom of rotations in the target space; so we can write $\chi^i = \chi(z) S^i (t, x, y)$. In~\cite{KurSh}, it was shown that the corresponding two-dimensional effective action on the domain wall is

\beq
\begin{gathered}
S_{eff} = \frac{1}{2}\,\frac{\chi_{0}^{2}}{m_{\varphi}}I_{1}\int \d t\,\d x\,\d y\,\left(\partial_{p}S^{i}\partial^{\s p}S^{i}\right)\quad,\\
\qquad S^{i}S^{i}=1\quad,\qquad  i=1,2,3 \quad,\qquad p=0,1,2\quad,
\end{gathered}
\label{O(3)Action}
\eeq
where
\beq 
I_{1}=\frac{m_{\varphi}}{\chi_{0}^{2}}\int\chi(z)^{2}\d z\quad.
\eeq

The topic of skyrmions on a domain wall has attracted a considerable attention~\cite{Nitta1,Nitta2,Nitta3,Nitta4,GudnasonNitta,KudPiZak,ZhouEzawa,Seibold,JenningsSutcliffe,Blyankinshtein,GiantEzawa,Volovik1,Volovik2}. It is important from both the theoretical viewpoint (see, for example, the interesting results reported in~\cite{WatanabeMurayama} on noncommuting momenta) and in relation to the active experimental research on the magnetic skyrmions in thin films and domain walls, and their possible practical applications for data storage and logic gates~\cite{RommingHanneken,RosslerBogdanov,ZhangEzawaZhou,Zhouetal,Benitezetal,HeinonenSomaily,BoulleVogel,Wiesendanger,NagaosaTokura,Kyoungetal}. The model described in Section~\ref{Moduli on a domain wall} with the Lagrangian ${\cal L}={\cal L}_0+{\cal L}_{\chi}$ supports skyrmions on domain walls within a setup very similar to the one proposed in the pioneering work~\cite{BelavinPolyakov}. Its main advantage is that it describes both the bulk and a domain wall with skyrmions within a unified approach wherein the properties of lower-dimensional topological
defects follow from the general description of the (3+1)-dimensional physics. The domain wall itself arises as a symmetry-breaking solution ensuing from the translationally invariant Lagrangian. The domain wall solution allows to derive the effective field theory on that wall from the initial model. The skyrmion, in its turn, emerges as a solution of that effective field theory. This way, we do not need to manually add any symmetry-breaking terms like the anisotropy term in~\cite{GiantEzawa} or the commonly employed Dzyaloshinskii-Moriya interaction. In the same way as for the domain line in the $\Z_2$ model of the previous section, the stability of such a skyrmion is of a topological nature.

It is convenient to visualize such a defect, by imagining a target space position as a 3-dimensional vector attached to each point of the domain wall plane, and having its  $x$, $y$ and $z$ components equal to $S^1$, $S^2$ and $S^3$, respectively. This representation becomes exact, when target space position represents a spin vector. Let us introduce spherical coordinates $(\theta,\psi)$ in the target space, and polar coordinates $(r,\phi)$ on the plane with $r = 0$ at the centre of the defect. For spin systems, the investigated configurations include the simplest case depicted in Figure~\ref{hedgehog}. There, the vector in the target space goes from the north pole in the centre of the defect to the south pole at infinity. The projection of this vector on the domain wall plain always points along the radius-vector. In a more complicated spiral case (Figure~\ref{spiral}), the projection of a target space vector on the plane winds around the centre as the value of $r$ increases.

Topologically, the defects we are looking for are easy to understand through the stereographic projection of the the real plane of domain wall on a sphere, where the centre of the defect corresponds to one pole and the vacuum at infinity -- to the other one. The topologically non-trivial mappings are possible between the real and target-space spheres which can be classified using the winding number
\beq
N = \frac{1}{8\pi} \int \varepsilon_{ij} \vec{\chi} (\partial^{\s i} \vec{\chi} \times 
\partial^{\s j} \vec{\chi}) \d x \d y\quad.
\label{VindingNumber}
\eeq

The simplest way to get an exact solution for $S^i$, as functions of the space coordinates, for a skyrmion relies on the fact that it is topologically identical to the instanton in ${1+1}$ dimensions, with a Euclidean time. Then, for a skyrmion with $N = 1$ which goes from the north pole at the centre to the south pole at infinity, we can immediately write the answer~\cite{BPST,NSVZ}:

\begin{equation}
\begin{alignedat}{9}
S^1 (\vec{x}) &= \frac{2(\vec{x} - \vec{x_0}) \cdot \vec{y}} {(\vec{x} - \vec{x_0})^2 + \vec{y}^{\, 2}}
\quad&&,\qquad
&&S^2 (\vec{x}) = \frac{2(\vec{x} - \vec{x_0}) \times \vec{y}} {(\vec{x} - \vec{x_0})^2 + \vec{y}^{\, 2}}\quad&&,
\\
S^3 (\vec{x}) &= \frac{\vec{y}^2 - (\vec{x} - \vec{x_0})^2} {(\vec{x} - \vec{x_0})^2 + \vec{y}^{\, 2}}
\quad&&,\qquad
&&(\vec{a} \times \vec{b} = a_1 b_2 - a_2 b_1)\quad&&.
\end{alignedat}
\label{SkyrmionSolution}
\end{equation}

Here $\vec{x}$ represents the coordinates in the plane, $\vec{x_0}$ is the the centre of the skyrmion, and vector $\vec{y}$ defines  the orientation and size $\vert\vec{y}\vert$ of the defect. These functions for $\vec{y} = (1, 0)$ are shown in Figure~\ref{SkyrmionImages} both as three different components of $S$ and as three-dimensional vectors on the domain wall plane. We see that $S^3$ decreases continuously and the regions with positive and negative $S^1$ and $S^2$ are symmetric with respect to $x$ and $y$ axes, respectively.

The skyrmion (\ref{SkyrmionSolution}) has four degrees of freedom: the two coordinates of the centre, the orientation, and the size. Considering them as functions of time while keeping the profile of the system fixed, we can derive an appropriate low-energy effective action in the same way as was done above for the one-dimensional domain wall. Since this was done for a very similar model in~\cite{Vlasii}, we refer the reader to it.

The skyrmion solution again gives us a convenient set-up to apply the method of symmetry analysis. Such a solution was obtained from the system~\cite{BPST,NSVZ}
\beq
\partial_{\mu} S^i = - \epsilon^{ijk} S^j \epsilon_{\mu \nu} \partial_{\nu} S^k
\label{SKYRRR}
\eeq
with the boundary conditions for which the vector in the target space goes from one pole at the centre of the defect to another pole at infinity.

The system~\eqref{SKYRRR} permits for the symmetry group generated by the following operators:
\begin{equation}\begin{alignedat}{8}
    X_1 &= \partial_x\quad,\qquad
    X_2 = \partial_y\quad,\qquad
    &&X_3 &&= y\jo\partial_x - x\jo\partial_y\quad,\\
    X_4 &= x\jo\partial_x + y\jo\partial_y\quad,\qquad
    &&X_5 &&= S^{\jo 1}\jo\partial_{S^{\jo 2}} -S^{\jo 2}\jo\partial_{S^{\jo 1}}\quad&&,\\
    X_6 &= S^{\jo 2}\jo\partial_{S^{\jo 3}} -S^{\jo 3}\jo\partial_{S^{\jo 2}}\quad,\qquad
    &&X_7 &&= S^{\jo 3}\jo\partial_{S^{\jo 1}} -S^{\jo 1}\jo\partial_{S^{\jo 3}}\quad&&.
\end{alignedat}\end{equation}
The first four are responsible for the translational, boost and scaling invariance in the coordinate space; the other three~---~for the rotational invariance in the target space, see the discussion in Appendix below.

\section{Conclusions}

After reviewing the construction of domain walls with orientational moduli localized on them, we compared the new results~\cite{GaniLizRad,GaniLizRad2} on analytic solutions for such a system with the numerical results obtained thitherto~\cite{KurSh}. Then, within a simplified model with two scalar fields possessing $\Z_2$ symmetries, we demonstrated the possibility of existence of domain lines localized on two-dimensional walls, and derived the corresponding effective action.
We studied skyrmions emerging within an effective $\mathbb{O}(3)$ sigma-model localized on a domain wall, and provided a solution analogous to the one known for instantons. We also analyzed the symmetries of the initial model which supports domain walls, as well as the symmetries of the effective low-energy theory of the domain wall world volume.

\section*{Acknowledgements}

We are grateful to Mikhail Shifman for helpful guidance and to Vakhid Gani for fruitful discussions.

\appendix
\gdef\thesection{Appendix \Alph{section}}
\section{Symmetry analysis}

We start out with addressing a general approach to eliminating \textit{all} continuous symmetries of a system of differential equations of the form
\begin{equation}
\begin{gathered}
    X = \xi^{\jo j}(x^{\jo k},\jo u^{\jo l})\, \partial_{x^{j}}
      + \eta^{\jo s}(x^{\jo k},\jo u^{\jo l})\jo \partial_{u^{s}} \quad,\\
    j,\jo k= 1,\jo2\jo \ldots,m \quad,\qquad l,\jo s = 1,\jo2,\jo \ldots,n \quad.  
    \label{X}
\end{gathered}
\end{equation}
Here $x^{\jo j}$ and $u^{\jo l}\jo(x^{\jo j})$ stand for independent variables and their functions, respectively.

The form of the symmetry operator~\eqref{X} indicates that we are seeking so-called \textit{point symmetries}~---~those for which the functions $\xi^{\jo j}$ and $\eta^{\jo l}$ depend only on the coordinates $x^{\jo j}$ and functions $u^{\jo l}\jo(x^{\jo j})$. In principle, nothing should prohibit us from allowing the functions $\xi^{\jo k}$ and $\eta^{\jo l}$ to depend on the derivatives of functions as well.\s\footnote{~Such symmetries (referred to as \textit{generalized}) are not  necessarily unphysical. A textbook example of this kind of symmetry having a physical meaning is the conservation of the Runge-Lenz vector. It can be demonstrated that the conservation of this vector stems from a generalized symmetry of the Kepler problem.
}
However, in the current paper we limit ourselves to point symmetries solely.

All the symmetries of a system of differential equations can be found with the aid of a general algorithm described in~\cite{Olver}. Being straightforward but quite lengthy, this algorithms is often implemented by computer packages. These are especially needed when the system is complicated, i.e., contain a large number of variables or/and dependent functions, or contains higher-order derivatives.
\footnote{~We recommend using the software that comes with the book~\cite{Lie}.}

Oftentimes, the knowledge of symmetries helps to perform integration of the system of differential equations. This task is more challenging than the search for symmetries, and each specific problem has to be treated uniquely. Still, various methods exist, of which the most universal one relies on the employment of \textit{differential invariants} of the symmetries, which are the differential manifolds on which the operator~\eqref{x} acts invariantly.

Below we remind the definition of a continuous symmetry for a given manifold, and extrapolate the construction to differential manifolds.

\subsection{Manifolds, invariants, symmetries}

In an ambient space of coordinates $x^i$, $i=1, \ldots,\jo m$, we define a manifold $\mathcal{M}$ as a set of all points satisfying a certain system of $s$ 
equations, $\,$with $\,s<m\,$:
\begin{alignat}{99}
    &\mathcal{M}\jo: \quad F^{\jo \sigma}\jo(x^{\jo j}) = 0 \quad&&,
    \qquad\sigma = 1,\jo 2,\jo \ldots\jo,\jo s\quad
    &&.
        \label{M}
\intertext{
A symmetry of the manifold is its property of staying invariant under the action of a differential operator $\,X\,$:}
    &X\left.F^{\jo \sigma}\jo(x^{\jo k}) \right|_{\mathcal{M}}= 0 \quad&&,\qquad\sigma = 1,\jo2,\jo \ldots\jo,\jo N \quad&&,
    \label{XM}
\intertext{where}
    &X = \xi^{\jo j}(x^{\jo k})\, \partial_{x^{\jo j}}
    \quad&&,\qquad j,\jo k = 1,\jo \jo\ldots\jo,\jo m \quad&&.
    \label{x}
\end{alignat}
The subscript $\,\mathcal{M}\,$ in the equation~\eqref{XM} serves to emphasize that the expression $\,X\,F^{\jo \sigma}\jo(x^{\jo k})\,$ vanishes only in the points of the manifold. 

Let $\,r\,$ stand for the rank of Wronskian of the system~\eqref{M}:
\begin{equation}
    r = \rank \left( \dfrac{\partial\jo F^{\jo \sigma}}{\partial\jo x^{\jo k}} \right) \quad. \label{XMM}
\end{equation}
When $\,r = s\,$, the system is said to have a full rank, while the manifold is termed regular. {To evaluate~\eqref{XM} in a point of the manifold, we solve~\eqref{M} for some $\,r\,$ coordinates, i.e., express them through the other $\,m-r\,$ coordinates. Insertion of the so-obtained $\,r\,$ expressions into~\eqref{XM} furnishes $\,X\,F^{\jo \sigma}\jo(x^{\jo k})\,$ as functions of the said $\,m-r\,$ coordinates.

As an example, one can consider the manifold $\mathcal{M}_0$ defined by the equation
\begin{equation}
 \mathcal{M}_0:\quad F_0\jo (x,\jo y\jo,z) =  
    \me^{\jo-\arctan\jo x/y} - \dfrac{x^{\jo 2}+y^{\jo 2}}{z} = 0 \quad.
\end{equation}
It stays invariant under the action of the operator
\begin{equation}
    X_0 = y\jo \partial_{x} -\jo x \jo \partial_{y} + z \jo \partial_z\quad,
\end{equation}
since
\begin{equation}
    X_0 \, F_0 = - F_0
    \quad.
\end{equation}

The equation
\begin{equation}
    X_0\,I\jo(x) = 0
\end{equation}
renders all the functionally-independent invariants of the operator $X_0$: 
\begin{equation}
    I_1 = x^{\jo 2}+y^{\jo 2} \quad, \quad I_2 = \log z - \arctan x/y \quad.
\end{equation}
Any other invariant of $X_0$ can be expressed as a function of those. For example, the manifold  $\mathcal{M}_0$ reads as:
\begin{equation}
    \mathcal{M}_0:\quad\me^{\jo I_2} - I_1 = 0\quad.
\end{equation}

\subsection{Symmetries of differential manifolds}

Let us now introduce, in addition to the independent variables $x^{\jo j}$, functions $\,u^{\jo l}\jo(x^{\jo j})\,$. Their partial derivatives will be denoted as:
\begin{equation}
    p^{\jo l}_j \equiv \dfrac{\partial\jo u^{\jo l}}{\partial x^{\jo j}}
    \quad,\qquad
    q^{\jo l}_{j\jo k} \equiv \dfrac{\partial\jo u^{\jo l}}{\partial x^{\jo j} \jo \partial x^{\jo k}}
    \quad, \qquad \ldots
    \label{parti}
\end{equation}
Then a system of differential equations for the functions $u^{\jo l}\jo(x^{\jo j})$ and their derivatives can be treated as a \textit{differential manifold}:
\begin{equation}\begin{gathered}
    \mathcal{M}\jo: \quad F^{\jo \sigma}\jo[x^{\jo j},\jo u^{\jo l},\jo
    p^{\jo l}_j,\jo q^{\jo l}_{j\jo k},\jo \ldots
    ] = 0 \quad,
    \qquad\sigma = 1,\jo2,\jo \ldots,N\quad.
        \label{diffman}
\end{gathered}\end{equation}

In order to calculate the action of the symmetry operator
\begin{equation}
    X = \xi^{\jo j}(x^{\jo k},\jo u^{\jo l})\, \partial_{x^{j}}
      + \eta^{\jo s}(x^{\jo k},\jo u^{\jo l})\jo \partial_{u^{s}}
    \label{XX}
\end{equation}
on the equations of the system~\eqref{diffman}, we have to define its action on the partial derivatives~\eqref{parti}. This requires the construction of the so-called \textit{prolongated operator}. When looking for the symmetries of the differential equation~\eqref{diffman} of the order $\varkappa$, one first has to construct the $\varkappa$-th \textit{prolongation} of the symmetry operator. Such prolongations have the form of
\begin{subequations}
\begin{alignat}{8}
    \underset{1}{X} &= X &&+ \zeta^{\jo l}_j\jo \partial_{p^{\jo l}_j} \quad&&,\\
    \underset{2}{X} &= \underset{1}{X} &&+ \zeta^{\jo l}_{j\jo k}\jo \partial_{q^{\jo l}_{j\jo k}} \quad&&,\\
    &\ldots &&  \quad&&.\nonumber
\end{alignat}
\end{subequations}
While the functions $\xi^{\jo j}$ and $\eta^{\jo s}$ in~\eqref{XX} are allowed to depend on $x^{\jo j}$ and $u^{\jo l}$ only, the functions $\zeta$ depend on the partial derivatives as well:
\begin{subequations}
\begin{alignat}{8}
    \zeta^{\jo l}_j &= \zeta^{\jo l}_j \jo &&(\xi^{\jo j},\jo \eta^{\jo s},\jo p^{\jo l}_j) \quad&&,\\
    \zeta^{\jo l}_{j\jo k} &= \zeta^{\jo l}_{j\jo k} \jo &&(\xi^{\jo j},\jo \eta^{\jo s},\jo p^{\jo l}_j,\jo q^{\jo l}_{j\jo k}) \quad&&,\\
    &\ldots &&  \quad&&.
\end{alignat}
\end{subequations}
More importantly, these functions are fully determined by the form of operator $X$: given the functions  $\xi^{\jo j}$ and $\eta^{\jo s}$, one can find the prolongated operator. For convenience of the further derivations, it will be convenient to define the covariant derivative as
\begin{equation}
    D_j = \partial_{x^{\jo j}} + p^{\jo l}_j \jo \partial_{u^{\jo l}} \quad.
\end{equation}
Then, the first and higher prolongations can be calculated using
\begin{subequations}
\begin{alignat}{7}
    \zeta^{\jo l}_j &= D_j \jo \eta^{\jo l} - p ^{\jo l}_s \jo D_j \jo \xi^{\jo s} \quad,\label{zeta1}\\
    \zeta^{l}_{i_1\jo i_2\jo \ldots i_k \jo j} &= D_j \jo \zeta^{\jo l}_{i_1\jo i_2\jo \ldots i_k} - p^{\jo l}_{i_1\jo i_2\jo \ldots i_k \jo s}\jo D_j \jo \xi^{\jo s} \quad.
\end{alignat}
\end{subequations}

The result of action of a prolongated operator on the differential equations of the system will have the form of a polynomial in variables $p^{\jo l}_j$,~$q^{\jo l}_{j\jo k}$,~$\ldots\;$. By setting all its coefficients equal to zero, one obtains a system of differential equations for the functions $\xi^{\jo j}$ and $\eta^{\jo s}$, which are called the \textit{determining equations}. In most cases, these equations are easy so solve, and the set of their solutions determines the full group of the continuous symmetries of the equation.

Let us briefly summarize the steps of the described procedure.
\begin{enumerate}
\item For a given set of independent variables and unknown functions, write down the most general form of the symmetry operator, prolongated up to the highest order of the derivative in the system.

\begin{sloppypar}
For instance, the for system~\eqref{SKYRRR}, the variables and functions are ${\{x^{\jo 1}\equiv x,\jo x^{\jo 2}\equiv y\}}$ and ${\{S^1,\jo S^2,\jo S^3\}}$, while the symmetry operator, prolongated once, has the form
\end{sloppypar}
\begin{equation}\begin{alignedat}{99}
    \underset{1}{X} =
    \tikzmark{bbb}
    &\xi^{\jo 1}\jo \partial_{x} &&+ \xi^{\jo 2}\jo \partial_y &&+ 
    \eta^{\jo 1}\jo \partial_{S^{\jo 1}} &&+
    \eta^{\jo 2}\jo \partial_{S^{\jo 2}} &&+
    \eta^{\jo 3}\jo \partial_{S^{\jo 3}}
    \tikzmark{eee}
    \\ +\:
    &\zeta^{\jo 1}_1\jo \partial_{p^{\jo 1}_1} &&+
    \zeta^{\jo 1}_2\jo \partial_{p^{\jo 1}_2} &&+
    \zeta^{\jo 2}_1\jo \partial_{p^{\jo 2}_1} &&+
    \zeta^{\jo 2}_2\jo \partial_{p^{\jo 2}_2} &&+
    \zeta^{\jo 3}_1\jo \partial_{p^{\jo 3}_1} +
    \zeta^{\jo 3}_2\jo \partial_{p^{\jo 3}_2}
    \quad.
\label{Xs}
\end{alignedat}
\InsertOverBrace[draw=black,text=black]{bbb}{eee}{$X$}
\end{equation}
With aid of~\eqref{zeta1}, the functions $\zeta^{\jo l}_j$ are found to be:
\begin{equation}\begin{alignedat}{9}
    \zeta^{\jo l}_j &= \eta^{\jo l}_{x^{\jo j}} &&+ p^{\jo 1}_{j}\jo \eta^{\jo l}_{S_{\jo 1}} &&+ p^{\jo 2}_{j}\jo \eta^{\jo l}_{S_{\jo 2}} &&+ p^{\jo 3}_{j}\jo \eta^{\jo l}_{S_{\jo 3}} \\
     &- (\xi^{\jo 1}_{x^{\jo j}} &&+ p^{\jo 1}_{j}\jo \xi^{\jo 1}_{S^{\jo 1}} &&+ p^{\jo 2}_{j}\jo \xi^{\jo 1}_{S^{\jo 2}} &&+ p^{\jo 3}_{j}\jo \xi^{\jo 1}_{S^{\jo 3}})\jo p^{\jo l}_1  \\
     &- (\xi^{\jo 2}_{x^{\jo j}} &&+ p^{\jo 1}_{j}\jo \xi^{\jo 2}_{S^{\jo 1}} &&+ p^{\jo 2}_{j}\jo \xi^{\jo 2}_{S^{\jo 2}} &&+ p^{\jo 3}_{j}\jo \xi^{\jo 2}_{S^{\jo 3}}) \jo p^{\jo l}_2 \quad.
\end{alignedat}\end{equation}

\item Act with the prolongated operator on each equation of the system. 

For skyrmions, the system reads as:
\begin{equation}
S^{\jo i} \jo p^{\jo}
    p^{\jo l}_j = -\epsilon^{\jo l \jo s \jo t}\jo
    \epsilon_{j\jo k} \jo S^{\jo s} \jo p^{\jo t}_{k}\quad.
\label{sksys}
\end{equation}
The result of action of the operator~\eqref{Xs} on the functions~\eqref{sksys} is straightforward to obtain, but is too long to present here.

\item \label{soso} Solve (algebraically) the original system of equations for any $\,N\,$ partial derivatives, and substitute these into the expression resulting from the action of the prolongated operator on the equations~\eqref{diffman}. See also the discussion after the equation~\eqref{XMM}.

\item Obtain the system of \textit{determining equations} by setting zero all the coefficients next to monomials in partial derivatives.\s\footnote{~By monomials in partial derivatives we mean expressions of the form $\,(p^{\jo l}_{i})^{\jo \alpha}\jo (q^{\jo s}_{j \jo k})^{\jo \beta}\jo \ldots$} Be mindful, that in the preceding step we have already restricted ourselves to the manifold.

\item Solve the determining equations. This step is usually very technical but straightforward. The more complicated the system of equations is and the less symmetries it respects, the more determining equations can be obtained, and the simpler they are.

Above we assumed the system of $N$ differential equations to be non-degenerate, which is often not the case. For example, of the six equations in the system~\eqref{sksys} not all are independent. One can solve four of them for some partial derivatives, and the other two equations will be satisfied automatically (if one keeps in mind that the functions $S^{\jo l}$ are defined on a unit sphere):
\begin{equation}\begin{alignedat}{9}
    p^{\jo 2}_1 &= \left((S^{\jo 1})^2 -1\right) \left(p^{\jo 1}_{1} \jo S^{\jo 1}\jo S^{\jo 2} + p^{\jo 1}_{2} \jo S^{\jo 3}
    \right) \quad&&,\\
    p^{\jo 2}_2 &= \left((S^{\jo 1})^2 -1\right) \left(p^{\jo 1}_{2} \jo S^{\jo 1}\jo S^{\jo 2} - p^{\jo 1}_{1} \jo S^{\jo 3}
    \right) \quad&&,\\
    p^{\jo 3}_1 &= \left((S^{\jo 1})^2 -1\right) \left(p^{\jo 1}_{1} \jo S^{\jo 1}\jo S^{\jo 3}-p^{\jo 1}_{2} \jo S^{\jo 2}\jo
    \right) \quad&&,\\
    p^{\jo 3}_2 &= \left((S^{\jo 1})^2 -1\right) \left(p^{\jo 1}_{1} \jo S^{\jo 2} + p^{\jo 1}_{2}\jo S^{\jo 1} \jo S^{\jo 3}
    \right) \quad&&.
    \label{cuddle}
\end{alignedat}\end{equation}

For this reason, one should only analyze four instead of six equations. Equations~\eqref{cuddle} can also serve for restricting to the manifold in the step~\ref{soso}.

\end{enumerate}


\begin{thebibliography} {99}
{\small

\bibitem{HananyTong}
A. Hanany, D. Tong, 
{\em Vortices, Instantons and Branes,} 
JHEP {\bf 0307}, 037 (2003) 
[arXiv:hep-th/0306150].

\bibitem{AuzziBolognesi}
R. Auzzi, S. Bolognesi, J. Evslin, K. Konishi, A. Yung, 
{\em Nonabelian Superconductors: Vortices and Confinement in ${\mathcal N}=2$ SQCD,} 
Nucl. Phys. B {\bf 673}, 187 (2003) 
[arXiv:hep-th/0307287].


\bibitem{ShifmanYung-ConfMon}
M. Shifman, A. Yung, 
{\em Non-Abelian String Junctions as Confined Monopoles,} 
Phys. Rev. D {\bf 70}, 045004 (2004) 
[arXiv:hep-th/0403149].


\bibitem{Witten} 
E.~Witten,
{\em Superconducting Strings,}
Nucl.\ Phys.\ B {\bf 249}, 557 (1985).


\bibitem{Sh1}
M.~Shifman, 
{\em Simple Models with Non-Abelian Moduli on Topological Defects,} 
Phys. Rev. D {\bf 87}, 025025 (2013) 
[arXiv:1212.4823 [hep-th]].


\bibitem{KurSh}
M.~Shifman, E.~Kurianovych,
{\em Non-Abelian Moduli on Domain Walls,} 
Int. J. Mod. Phys. A {\bf 29}, 1450193 (2014)
[arXiv:1407.7144 [hep-th]].


\bibitem{GaniLizRad}
V. A. Gani, M. A. Lizunova, R. V. Radomskiy,
{\em Scalar triplet on a domain wall: an exact solution,} 
JHEP {\bf 04}, 043 (2016)
[arXiv:1601.07954 [hep-th]].


\bibitem{GaniLizRad2}
V. A. Gani, M. A. Lizunova, R. V. Radomskiy,
{\em Scalar triplet on a domain wall,} 
J. Phys.: Conf. Ser. {\bf 675}, 012020 (2016)
[arXiv:1602.04446 [hep-th]].


\bibitem{MShYu}
S. Monin, M.~Shifman, A. Yung,
{\em Calculating Extra (Quasi)Moduli on the Abrikosov-Nielsen-Olesen string with Spin-Orbit Interaction,}
Phys. Rev. D {\bf 88}, 025011 (2013) 
[arXiv:1305.7292 [hep-th]].


\bibitem{Rajaraman}
R. Rajaraman,
{\em Solitons and Instantons}, (Elsevier, Amsterdam, 1996).

\bibitem{Auzzi}
R. Auzzi, M. Shifman, A. Yung,
{\em Domain Lines as Fractional Strings,}
Phys. Rev. D {\bf 74}, 045007 (2006) 
[arXiv:hep-th/0606060].


\bibitem{ShBook}
M. Shifman,
{\em Advanced Topics in Quantum Field Theory}, (Cambridge University Press, 2012).

\bibitem{Nitta1}
M. Nitta,
{\em Correspondence between Skyrmions in 2+1 and 3+1 Dimensions,}
Phys. Rev. D {\bf 87}, 025013 (2013)
[arXiv:1210.2233 [hep-th]].


\bibitem{Nitta2}
M. Nitta,
{\em Matryoshka Skyrmions,}
Nucl. Phys. B {\bf 872}, 1 (2013) 
[arXiv:1211.4916 [hep-th]].


\bibitem{Nitta3}
M. Nitta,
{\em Josephson vortices and the Atiyah-Manton construction,}
Phys. Rev. D {\bf 86}, 125004 (2012)
[arXiv:1207.6958 [hep-th]].


\bibitem{Nitta4}
M. Nitta,
{\em Josephson instantons and Josephson monopoles in a non-Abelian Josephson junction,}
Phys. Rev. D {\bf 92}, 045010 (2015)
[arXiv:1503.02060 [hep-th]].


\bibitem{GudnasonNitta}
S. B. Gudnason, M. Nitta,
{\em Domain wall Skyrmions,}
Phys. Rev. D {\bf 89}, 085022 (2014) 
[arXiv:1403.1245 [hep-th]].


\bibitem{KudPiZak}
A. Kudryavtsev, B. Piette, W.J. Zakrzewski,
{\em Skyrmions and domain walls in (2+1) dimensions,}
Nonlinearity, {\bf 11}, 4 (1998) 
[arXiv:hep-th/9709187].


\bibitem{ZhouEzawa}
Yan Zhou, Motohiko Ezawa,
{\em A reversible conversion between a skyrmion and a domain-wall pair in junction geometry,}
Nature Communications {\bf 5}, 4652 (2014) 
[arXiv:1404.3350 [cond-mat.str-el]].


\bibitem{Seibold}
G. Seibold,
{\em Vortex, skyrmion and elliptical domain wall textures in the two-dimensional Hubbard model,}
Phys. Rev. B {\bf 58}, 15520 (1998) 
[arXiv:cond-mat/9809113 [cond-mat.str-el]].


\bibitem{JenningsSutcliffe}
P. Jennings, P. Sutcliffe,
{\em The dynamics of domain wall Skyrmions,}
Journal of Physics A {\bf 46}, 465401 (2013) 
[arXiv:1305.2869 [hep-th]].


\bibitem{Blyankinshtein}
N. Blyankinshtein,
{\em Q-lumps on Domain Wall with Spin-Orbit Interaction,}
Phys. Rev. D {\bf 93}, 065030 (2016) 
[arXiv:1510.07935 [hep-th]].


\bibitem{GiantEzawa}
M. Ezawa,
{\em Giant Skyrmions Stabilized by Dipole-Dipole Interactions in Thin Ferromagnetic Films,}
Phys. Rev. Lett. {\bf 105}, 197202 (2010) 
[arXiv:1007.4048 [cond-mat.str-el]].


\bibitem{Volovik1}
U. Parts, E. V. Thuneberg, G. E. Volovik, J. H. Koivuniemi, V. M. H. Ruutu, M. Heinila, J. M. Karimäki, M. Krusius,
{\em Vortex sheet in rotating superfluid $^{3}\mathit{A}$,}
Phys. Rev. Lett. {\bf 72}, 3839 (1994). 

\bibitem{Volovik2}
M.T. Heinila, G.E. Volovik,
{\em Bifurcations in the growth process of a vortex sheet in rotating superfluid,}
Physica B: Condensed Matter {\bf 210}, 300 (1995). 

\bibitem{WatanabeMurayama}
H. Watanabe, H. Murayama,
{\em Noncommuting Momenta of Topological Solitons,}
Phys. Rev. Lett. {\bf 112}, 191804 (2014) 
[arXiv:1401.8139 [hep-th]].


\bibitem{RommingHanneken} 
N. Romming, C. Hanneken, M. Menzel, J. E. Bickel, B. Wolter, K. von Bergmann, A. Kubetzka, R. Wiesendanger,
{\em Writing and Deleting Single Magnetic Skyrmions,}
Science {\bf 341}, 6146 (2013).


\bibitem{RosslerBogdanov} 
U. K. Rossler, A. N. Bogdanov, C. Pfleiderer,
{\em Spontaneous skyrmion ground states in magnetic metals,}
Nature {\bf 442}, 797-801 (2006).


\bibitem{ZhangEzawaZhou} 
X. Zhang, M. Ezawa, Y. Zhou,
{\em Magnetic skyrmion logic gates: conversion, duplication and merging of skyrmions,}
Scientific Reports {\bf 5}, 9400 (2015) [arXiv:1410.3086 [cond-mat.mes-hall]].


\bibitem{Zhouetal} 
Y. Zhou, E. Iacocca, A. Awad, R. K. Dumas, F. C. Zhang,	H. B. Braun, J. Akerman,
{\em Dynamically stabilized magnetic skyrmions,}
Nature Communications {\bf 6}, 8193 (2015).


\bibitem{Benitezetal} 
M. J. Benitez, A. Hrabec, A. P. Mihai, T. A. Moore, G. Burnell, D. McGrouther, C. H. Marrows, S. McVitie,
{\em Magnetic microscopy and topological stability of homochiral Neel domain walls in a Pt/Co/AlOx trilayer,}
Nature Communications {\bf 6}, 8957 (2015).


\bibitem{HeinonenSomaily}
O. Heinonen, W. Jiang, H. Somaily, S. G. E. te Velthuis, A. Hoffmann,
{\em Generation of magnetic skyrmion bubbles by inhomogeneous spin Hall currents,}
Phys. Rev. B. {\bf 93}, 094407 (2016) 
[arXiv:1511.04630 [cond-mat.mes-hall]].


\bibitem{BoulleVogel} 
O. Boulle, J. Vogel, H. Yang, S. Pizzini, D. de Souza Chaves, A. Locatelli, T. Onur Mentes, A. Sala, L. D. Buda-Prejbeanu, O. Klein, M. Belmeguenai, Y. Roussigné, A. Stashkevich,	S. 
Mourad Chérif, L. Aballe, M. Foerster, M. Chshiev, S. Auffret, I. M. Miron, Gilles Gaudin,
{\em Room-temperature chiral magnetic skyrmions in ultrathin magnetic nanostructures,}
Nature Nanotechnology {\bf 11}, 449–454 (2016).


\bibitem{Wiesendanger} 
R. Wiesendanger,
{\em Nanoscale magnetic skyrmions in metallic films and multilayers: a new twist for spintronics,}
Nature Reviews Materials{\bf 1}, 16044 (2016).


\bibitem{NagaosaTokura} 
N. Nagaosa, Y. Tokura,
{\em Topological properties and dynamics of magnetic skyrmions,}
Nature Nanotechnology {\bf 8}, 899–911 (2013).


\bibitem{Kyoungetal} 
Kyoung-Woong Moon, Duck-Ho Kim, Sang-Cheol Yoo, Soong-Geun Je, Byong Sun Chun, Wondong Kim, Byoung-Chul Min, Chanyong Hwangб Sug-Bong Choe,
{\em Magnetic bubblecade memory based on chiral domain walls,}
Scientific Reports {\bf 5}, 9166 (2015).


\bibitem{BelavinPolyakov}
A.~M.~Polyakov, A.~A.~Belavin,
{\em Metastable States of Two-Dimensional Isotropic Ferromagnets,}
  JETP Lett.\  {\bf 22} (1975) 245
   [Pisma Zh.\ Eksp.\ Teor.\ Fiz.\  {\bf 22} (1975) 503].

\bibitem{BPST}
A. A. Belavin, A. M. Polyakov, A. S. Schwartz, Yu. S. Tyupkin,
{\em Pseudoparticle solutions of the Yang-Mills equations,}
Phys. Lett. B  {\bf 59} (1), 85 (1975).


\bibitem{NSVZ}
V. A. Novikov, M. A. Shifman, A. I. Vainshtein, V. I. Zakharov,
{\em Two-dimensional sigma models: Modelling non-perturbative effects in quantum chromodynamics,}
Phys. Rept. {\bf 116}, 103 (1984).

  
\bibitem{Vlasii}
N. D. Vlasii, C. P. Hofmann, F.-J. Jiang, U.-J. Wiese,
{\em Symmetry Analysis of Holes Localized on a Skyrmion in a Doped Antiferromagnet,}
Phys. Rev. B {\bf 86}, 155113 (2012) 
[arXiv:1205.3677 [cond-mat.str-el]].
\bibitem{Olver}

P Olver 2000 Applications of Lie Groups to Differential Equations, \textit{Springer}, 2nd edition.

\bibitem{Lie}

B J Cantwell 2002 Introduction to Symmetry Analysis, \textit{Cambridge University Press}.



}
\end{thebibliography}
\end{document}